\documentclass[runningheads]{llncs}
\usepackage{graphicx}
\usepackage{amsmath,amsfonts,amssymb}
\usepackage{subcaption}
\usepackage{bm}
 \usepackage[colorlinks=true,
linkcolor=blue,
citecolor=blue,
urlcolor=blue,
filecolor=blue,
pdfpagemode=None,
pdfstartview=FitH]{hyperref}
\usepackage{siunitx}
\newcommand{\mvect}[1]{\mathbf{#1}}

\newcommand{\trr}[1]{{#1}^{\!\top}}
\newcommand{\pd}[2]{\frac{\partial #1}{\partial #2}}

\newenvironment{rcases}
{\left.\begin{aligned}}
	{\end{aligned}\right\rbrace}
\usepackage{cite}
\begin{document}
\title{Towards  topology optimization of pressure-driven soft robots}
%
%

\author{Prabhat Kumar}
\authorrunning{P. Kumar}
%
\institute{Department of Mechanical and Aerospace Engineering, Indian Institute of Technology Hyderabad,  Telangana 502285, India\\
\email{pkumar@mae.iith.ac.in}}
\vspace{4mm}

\maketitle              

Published\footnote{This pdf is the personal version of an article whose final publication is available at \href{https://link.springer.com/chapter/10.1007/978-3-031-20353-4_2}{Microactuators, Microsensors and Micromechanisms}}\,\,\,in \textit{Microactuators, Microsensors and Micromechanisms:MAMM 2022}

\vspace{3mm}
\rule{\linewidth}{.15mm}
\begin{abstract}
Soft robots are made of compliant materials that perform their tasks by deriving motion from elastic deformations. They are used in various applications, e.g.,  for handling fragile objects, navigating sensitive/complex environments, etc., and are typically actuated by Pneumatic/hydraulic loads. Though demands for soft robots are continuously increasing in various engineering sectors, due to the lack of systematic approaches, they are primarily  designed manually. This paper presents a systematic density-based topology optimization approach to designing soft robots while considering the design-dependent behavior of the actuating loads. We use the Darcy law with the conceptualized drainage term to model the design-dependent nature of the applied pressure loads. The standard finite element is employed to evaluate the consistent nodal loads from the obtained pressure field. The robust topology optimization formulation is used with the multi-criteria objective. The success of the presented approach is demonstrated by designing a member/soft robot of the pneumatic networks (PneuNets). The optimized member is combined in several series to get different PneuNets. Their CAD models are generated, and they are studied with high-pressure loads in a commercial software. Depending upon the number of members in the PneuNets, different output motions are noted. 


\rule{\linewidth}{.15mm}
\end{abstract}
\section{Introduction}
\begin{figure}[h!]
	\centering
		\includegraphics[scale=1]{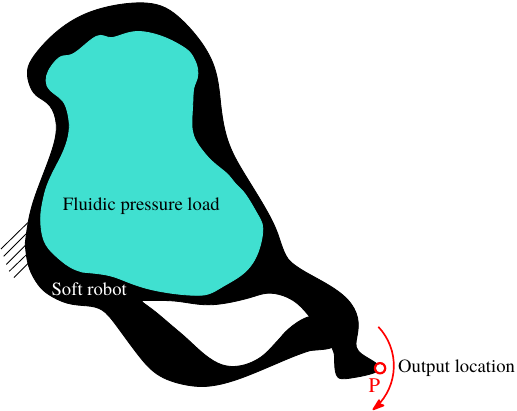}
		\caption{A schematic diagram for a soft robot.}	\label{fig:Schematic}
\end{figure}

Soft robots are constituted by compliant materials and have monolithic lightweight designs \cite{xavier2022soft}. Such robots are actuated primarily by pneumatic/hydraulic (fluidic pressure) loads and use motion obtained from elastic deformation to perform their tasks. Nowadays, they are being used in a wide range of applications, e.g., to handle fragile objects, fruits, and vegetables, in sensitive and unstructured environments for navigation, etc. In addition, they provide high power-to-weight ratios and help achieve complex motions \cite{xavier2022soft}. Therefore, interest in designing them for different applications is constantly growing. The pneumatically/hydraulically (pressure loads, air/water) actuated soft robots are sought the most and are used relatively more. In general, soft robots are designed manually using heuristic methods because of the lack of systematic approaches. Heuristic methods greatly depend upon the designers' knowledge and experience and may require many resources/iterations. Therefore, the goal of this paper is to present a systematic approach using topology optimization for designing pressure-driven soft robots. Figure~\ref{fig:Schematic} displays a schematic diagram of a soft robot with a bellow-shaped pressure loading chamber. When pneumatic/hydraulic loads inflate the chamber, it is desired that the output point P moves in a  bending motion, as shown by the red curved arrow. 

Topology optimization (TO) provides an optimized material distribution for a design problem by extremizing the desired objective under the given physical and geometrical constraints. In a typical  TO setting, the design domain is described by finite elements (FEs, cf.\cite{kumar2022honeytop90}), and each element is assigned a design variable $\rho \in[0,\,1]$. $\rho = 0$ and $\rho = 1$ indicate an element's void and solid states, respectively. The applied loads/actuating forces depending upon the applications soft robots are designed for, can be either design-dependent, e.g., fluidic pressure load \cite{kumar2020topology}, or constant. A design-dependent load changes its location, magnitude, and direction as TO advances and thus, poses several distinctive challenges \cite{kumar2020topology}.

Hiller and Lipson \cite{hiller2011automatic} used the evolutionary topology optimization approach to design soft robots. Chen et al.~\cite{chen2018topology} developed a soft cable-driven gripper using the level-set topology optimization method. Zhang et al. \cite{zhang2018design} developed a  soft pneumatic gripper by maximizing the output displacement. Pressure loads always act normal to the boundaries of the design domain, which alter with TO iterations. Therefore, including design-dependent behavior within the optimization formulation is essential for fluid pressure loads, which is not considered in Refs.~\cite{hiller2011automatic,chen2018topology,zhang2018design}. In addition, Refs. \cite{yin2001topology,sigmund1997design} mentioned that compliant mechanisms (CMs) optimized using TO suffer from point (single-node) connections, and thus, they either become challenging to realize or require post-processing. Further, the performance of the post-processed designs may not be the same as that of their numerical counterparts. Herein, we use the robust formulation presented in \cite{wang2011projection} to circumvent this issue, considering the design-dependent characteristics of pressure loads for designing soft robots. Readers can refer to Refs. \cite{kumar2019compliant,zhu2020design,kumar2021topology,kumar2020topology2} and references therein for designing various CMs for different applications using TO.
 
Hammer and Olhoff \cite{Hammer2000} were the first to consider the design-dependent nature of the pressure loads in TO while designing structures. Chen et al.~\cite{chen2001advances} used a fictitious thermal model to design pressure-actuated compliant mechanisms. Sigmund and Clausen \cite{Sigmund2007} presented the mixed-finite element-based approach. Panganiban et al.~\cite{Panganiban2010} employed a non-confirming FE method in their approach for pressure-actuated CMs. The solid isotropic material with penalization (SIMP) and moving isosurface threshold schemes are used by Vasista and Tong~\cite{vasista2012design}. de Souza and Silva \cite{de2020topology} used the method presented in Ref.~\cite{Sigmund2007} with a projection filter. Kumar et al.~\cite{kumar2020topology} presented a novel approach using the Darcy law for pressure field modeling. The method uses the standard FE method and works fine for designing 3D CMs \cite{kumar2020topology3Dpressure}. Thus, we adopt the method to model the pressure load. The prime goal herein is to design a member (soft robot) of the pneumatic network (PneuNets, cf. \cite{shepherd2011multigait}) to achieve the specified motion. The optimized member is further connected in several series to get different output motions with high-pressure loads. 

The remainder of the paper is structured as follows. Sec.~\ref{Sec:PressureLoadModeling} summarizes pressure load modeling using the Darcy law in brief. TO formulation is provided in Sec.~\ref{Sec:TopOptformulation}. Sec.~\ref{Sec:NumRD} presents an optimized design for a member of the PneuNet soft robot. The optimized design is extracted, and its CAD model is made. The different networks are generated from the CAD model and further analyzed with higher pressure loads in commercial software to achieve complex motions. Lastly, the conclusions are drawn in Sec.~\ref{Sec:Closure}.

\section{Pressure load modeling}\label{Sec:PressureLoadModeling}

 In this section, we present the pressure load modeling using the Darcy law in brief herein for completeness. One can refer to~\cite{kumar2020topology} for a detailed description.
 
  As TO advances, the material states of the associated FEs evolve. We already have given boundaries with input pressure and zero pressure loads at the initial stage of TO, i.e., a pressure difference across the domain is known. Therefore, using the Darcy low to determine the pressure field while assuming elements as porous medium is natural. Given Darcy law, the flux $\bm{q}$ is determined as
\begin{equation}\label{Eq:Darcyflux}
\bm{q} = -\frac{\kappa}{\mu}\nabla p = -K(\bar{\rho}) \nabla p,
\end{equation}
where $\nabla p$ is the pressure gradient. $\kappa$ and $\mu$ represent the permeability of the medium and the fluid viscosity, respectively. $\bar{\rho}$ indicates the physical design variable. $K(\bar{\rho})$ is called the flow coefficient. For element $e$, the flow coefficient is defined as \cite{kumar2020topology}
\begin{equation}\label{Eq:Flowcoefficient}
K(\bar{\rho_e}) = K_v\left(1-(1-\epsilon) \mathcal{H}(\bar{{\rho_e}},\,\beta_\kappa,\,\eta_\kappa)\right),
\end{equation}
where $\mathcal{H}(\bar{{\rho_e}},\,\beta_\kappa,\,\eta_\kappa) = \frac{\tanh{\left(\beta_\kappa\eta_\kappa\right)}+\tanh{\left(\beta_\kappa(\bar{\rho_e} - \eta_\kappa)\right)}}{\tanh{\left(\beta_\kappa \eta_\kappa\right)}+\tanh{\left(\beta_\kappa(1 - \eta_\kappa)\right)}}$, and $\epsilon=\frac{K_s}{K_v}$ is called flow contrast \cite{kumar2020topology3Dpressure}.  $K_s$ and $K_v$ are the flow coefficient of solid and void phases of element~$e$ respectively. $\left\{ \eta_\kappa,\,\beta_\kappa\right\}$ are termed flow parameters which define respectively the step position and slope of $K(\bar{\rho_e})$. To get the meaningful pressure distribution in a TO setting, drainage,  $Q_\text{drain}$, is conceptualized \cite{kumar2020topology,kumar2020topology3Dpressure}. The balanced equation of Eq.~\ref{Eq:Darcyflux} with the drainage  can be determined as \cite{kumar2020topology}:
\begin{equation}\label{Eq:stateequation}
\nabla\cdot\bm{q} -Q_\text{drain} = 0.
\end{equation} 
where ${Q}_\text{drain} = -D(\bar{\rho_e}) (p - p_{\text{ext}})$ with $D(\bar{\rho_e}) =  D_{\text{s}}\mathcal{H}(\bar{{\rho_e}},\,\beta_d,\,\eta_d)$. $\left\{\eta_\text{d},\,\beta_\text{d}\right\}$ are the drainage parameters. In view of the fundamentals of the finite element formulations, Eq.~\ref{Eq:stateequation} transpires to~\cite{kumar2020topology}
\begin{equation}\label{Eq:FinalbalanceEqu}
\mvect{Ap} = \mvect{0},
\end{equation}
in case both external pressure load and surface flux are set equal to zero, which is the case herein considered. $\mvect{A}$ and $\mvect{p}$ are the global flow matrix and pressure vector, respectively. Eq.~\ref{Eq:FinalbalanceEqu} gives the pressure field distribution within the design domain as TO advances. We find the consistent nodal loads from the pressure field distribution as~\cite{kumar2020topology}
\begin{equation}\label{Eq:nodalforce}
\mvect{F} = -\mvect{T}\mvect{p}
\end{equation}
where $\mvect{F}$ is the global force vector, and $\mvect{T}$ is the transformation matrix \cite{kumar2020topology}. To summarize, with TO iterations, Eq.~\ref{Eq:FinalbalanceEqu} and Eq.~\ref{Eq:nodalforce} are used to determine the pressure field and corresponding nodal force vector. 

\section{Topology optimization formulation}\label{Sec:TopOptformulation}
We use the robust formulation \cite{wang2011projection}, i.e., the eroded, intermediate, and dilated descriptions of the design field, to find the optimized designs for the soft robots. The worst objective of the eroded, intermediate, and dilated designs is minimized with the given volume fraction. Mathematically, the optimization formulation can be written as~\cite{kumar2022topological}
\begin{equation}\label{Eq:actualoptimization}
	\begin{rcases}
		\begin{aligned}
			&\underset{\bm{\rho}}{\text{min}}:\text{max}
		: \left(f_0(\bar{\bm{\rho}}^d),\,f_0(\bar{\bm{\rho}}^i),\,f_0(\bar{\bm{\rho}}^e)\right) \\
		&{\text{Subjected to}} :\\
		&\bm{\lambda}_m^1: \quad\quad \mathbf{A}_m\mathbf{p}_m|_{m=d,i,e} = \mathbf{0},\\
		&\bm{\lambda}_m^2: \quad\quad\mathbf{K}_m\mathbf{u}_m = \mathbf{F}_m = -\mathbf{T}_m \mathbf{p}_m\\
		&\bm{\lambda}_m^3: \quad\quad\mathbf{K}_m\mathbf{v}_m = \mathbf{F}_\mathrm{d}\\
		&\mu: \quad \quad V(\bar{\bm{\rho}}^d(\bm{\rho}))-V_d^*\le 0 \\
		&\quad\mvect{0}\le\bm{\rho}\le\mvect{1}\\
		&\text{Data:\,} V_d^*,\Delta\eta,r_\text{fill},P_\text{in}, \mathbf{F}_\mathrm{d}
		\end{aligned}
	\end{rcases},
\end{equation}
where $f_0$ is a multi-criteria objective, which is defined by $-s\frac{MSE}{SE}$ \cite{saxena2000optimal}. $MSE$ indicates the mutual strain energy, and $SE$ represents the strain energy. $MSE =\trr{\mathbf{v}}\mvect{Ku}$ and $SE = \frac{1}{2}\trr{\mvect{u}}\mvect{Ku}$. $s$ is a consistent scale factor. As per \cite{wang2011projection}, the volume fraction is applied using the dilated field and is updated with TO iterations. $\mathbf{u}$ and $\mathbf{v}$ are the global displacement vectors obtained in response to the actual and dummy forces, respectively. $\mathbf{K}$ is the global stiffness matrix. $\bar{\bm{\rho}}$ is the physical design vector.  $\bar{\rho_j}$ of element $j$ is determined as \cite{wang2011projection}
\begin{equation}\label{Eq:projectionfilter}
	\bar{\rho_j}(\tilde{\rho_j},\,\beta,\,\eta)  = \frac{\tanh{\left(\beta \eta\right)} + \tanh{\left(\beta (\tilde{\rho_j}-\eta)\right)}}{\tanh{\left(\beta \eta\right)} + \tanh{\left(\beta (1-\eta)\right)}},
\end{equation}
where $\eta\in[0,\,1]$ defines the threshold, and  $\beta\in[0,\,\infty)$ controls the steepness of the projection function. Generally, $\beta$ is increased from 1 to a large finite value using a continuation scheme. $\tilde{\rho_j}$, the filtered design variable of element $j$, is defined as~\cite{bruns2001topology}
\begin{equation}\label{Eq:densityfilter}
	\tilde{\rho_j} = \frac{\sum_{k=1}^{N_e} v_k \rho_k w(\mvect{x}_k)}{\sum_{k=1}^{N_e} v_k w(\mvect{x}_k)} 
\end{equation}
where $N_e$ indicates the total number of elements used to parameterize the design domain, and $v_k$ is the volume of neighboring element~$k$. $w(\mvect{x}_k)= \max\left(0,\,1-\frac{d}{r_\text{fill}}\right)$, is the weight function, wherein $d = ||\mvect{x}_j -\mvect{x}_k||$ is a Euclidean  distance between centroids $\mvect{x}_j$ and $\mvect{x}_k$ of elements  $j$ and $k$, respectively. One finds the derivative of $\bar{\rho_j}$ (Eq.~\ref{Eq:projectionfilter}) with respect to $\tilde{\rho_j}$ as:
\begin{equation}\label{Eq:derivativeprojectionfilter}
	\pd{\bar{\rho_j}}{\tilde{\rho_j}} = \beta\frac{1-\tanh(\beta(\tilde{\rho_j} -\eta))^2}{\tanh{\left(\beta \eta\right)} + \tanh{\left(\beta (1-\eta)\right)}}.
\end{equation}
Likewise, the derivative of $\tilde{\rho_j}$ (Eq.~\ref{Eq:densityfilter}) with respect to $\rho_k$  can be evaluated as
\begin{equation}\label{Eq:derivativefilteractual}
	\pd{\tilde{\rho_j}}{\rho_k} = \frac{v_k w(\mvect{x}_k)}{\sum_{i=1}^{N_e}v_i w(\mvect{x}_i)}. 
\end{equation}
Finally, using the chain rule one can find the derivative a function $f$ with respect to  $\rho_k$ as
\begin{equation}\label{Eq:ChainRule}
	\pd{f}{\rho_k} = \sum_{j= 1}^{Ne}\pd{f}{\bar{\rho_j}}\pd{\bar{\rho_j}}{\tilde{\rho_j}}\pd{\tilde{\rho_j}}{\rho_k},
\end{equation}

We use the modified SIMP material scheme \cite{sigmund2007morphology} for finding  Young's modulus of element $j$ as
\begin{equation}\label{Eq:SIMPModel}
	E_j= E_0 + (\bar{\rho_j})^\chi(E_1 -E_0), \qquad \bar{\rho_j}\in [0,\,1]
\end{equation}
where $\chi$ is the SIMP parameter. $E_0$ and $E_1$ are Youngs' moduli of the void and solid states of an FE, respectively. $\frac{E_0}{E_1} = \si{1e^{-6}}$ is set in this paper.

The Method of Moving Asymptotes (MMA, cf. \cite{svanberg1987method}), a gradient-based optimizer, is used to solve the optimization problem (Eq.~\ref{Eq:actualoptimization}). Thus, we need the sensitivities of the objective and constraint with respect to the design variables, which  are determined using the adjoint-variable method. Complete detail on sensitivity analysis can be found in \cite{kumar2020topology,kumar2022topological}.

\section{Numerical results and discussions}\label{Sec:NumRD}
\begin{figure}[h!]
	\centering
	\includegraphics[scale=1.25]{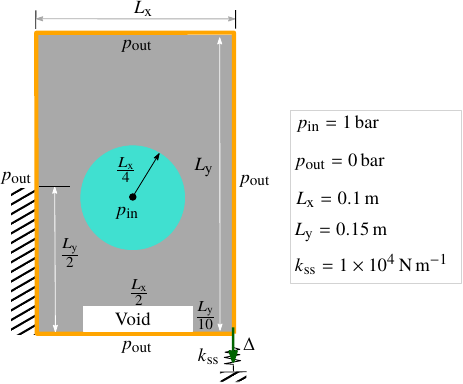}
	\caption{Design domain of a pressure-driven soft robot. $L_x =\SI{0.1}{\meter}$ and  $L_y = \SI{0.15}{\meter}$ are set, where $L_x$ and $L_y$ represent dimensions in $x-$ and $y-$directions, respectively.  The input pressure load of $\SI{1}{\bar}$ is applied using a circular chamber of radius $\frac{L_x}{4}$, as shown in the figure. The desired output motion is indicated by a green arrow. $k_\text{ss}$ indicates the work-piece stiffness. A predefined void region of dimension $\frac{L_x}{2}\times \frac{L_y}{10}$ is shown within the domain.} \label{fig:DesignDomain}
\end{figure}
In this section, a member of PneuNets  (Fig.~1(B) of \cite{shepherd2011multigait}) is designed using the presented method. Different PneuNets are constructed using the optimized design to achieve complex motions with high-pressure loads, which we analyze in a commercial software. 

The design domain is displayed in Fig.~\ref{fig:DesignDomain}. $L_x = \SI{0.1}{\meter}$ and $L_y = \SI{0.15}{\meter}$ indicate respectively the dimension in $x-$ and $y-$directions. The out-of-plane thickness is set to $\SI{0.001}{\meter}$. The input pressure load of $\SI{1}{\bar}$ is applied at the center of the domain, as shown in Fig.~\ref{fig:DesignDomain}. Edges of the domain experience zero pressure load. Half the length of the left edge is fixed (Fig.~\ref{fig:DesignDomain}). Workpiece of stiffness $k_{ss} = \SI{1e4}{\newton \per \meter}$ is used. The domain is parameterized using quadrilateral FEs herein.  $N_\text{ex}\times N_\text{ey} = 100 \times 150$ bi-linear quadrilateral FEs are used to describe the design domain, where FEs in $x-$ and $y-$directions are represented via $N_\text{ex}$ and $N_\text{ey}$, respectively. Each element is assigned a design variable that is considered constant within the element. The external move limit for the MMA optimizer is set to 0.1. The maximum number of the MMA iterations is fixed to 400. The filter radius $r_\text{fill} = 6.0 \times \min \left(\frac{L_x}{N_\text{ex}},\, \frac{L_y}{N_\text{ey}}\right)$ is set. A void region of dimension $\frac{L_x}{2}\times \frac{L_y}{10}$ exists within the domain, as displayed in Fig.~\ref{fig:DesignDomain}. We consider the plane stress and small deformation finite element formulation assumptions. The SIMP parameter $\chi =3$ is taken.
\begin{figure}[h!]
	\begin{subfigure}[t]{0.30\textwidth}
		\centering
		\includegraphics[scale=0.5]{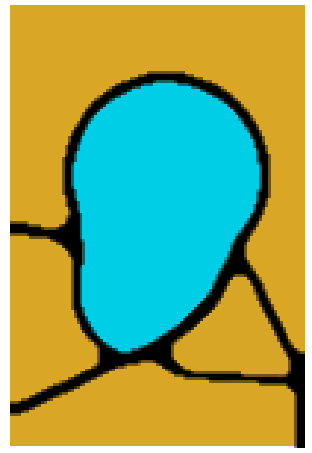}
		\caption{$\Delta = \SI{-0.231}{\milli \meter}$}
		$M_\text{nd}=0.42\%$,\,$V_f=0.135$\label{fig:erodedpnet}
	\end{subfigure}
	\quad
	\begin{subfigure}[t]{0.30\textwidth}
		\centering
		\includegraphics[scale=.5]{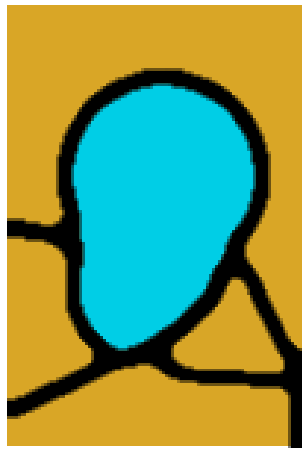}
		\caption{$\Delta = \SI{-0.224}{\milli \meter}$}
		$M_\text{nd}=0.38\%$,\,$V_f=0.2$\label{fig:intermediatepnet}
	\end{subfigure}
	\quad
	\begin{subfigure}[t]{0.30\textwidth}
		\centering
		\includegraphics[scale=.5]{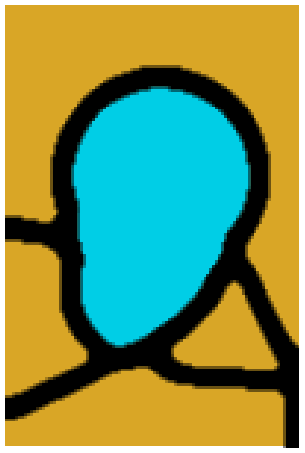}
		\caption{$\Delta = \SI{-0.209}{\milli \meter}$}
		$M_\text{nd}=0.76\%$,\,$V_f=0.26$\label{fig:dilatedpnet}
	\end{subfigure}
	\caption{Optimized design of a member of the PneuNet. Filter radius $r_\text{fill} = 6.0h$ is employed. $h=\min\left(\frac{L_x}{N_\text{ex}},\,\frac{L_y}{N_\text{ey}}\right)$. (\subref{fig:erodedpnet}) Eroded design, (\subref{fig:intermediatepnet}) Intermediate design, and (\subref{fig:dilatedpnet}) Dilated design.} \label{fig:solution}
\end{figure}

\begin{figure}[h!]
	\begin{subfigure}[t]{0.27\textwidth}
		\centering
		\includegraphics[scale=0.5]{Rv1i}
		\caption{}
		\label{fig:actualR1}
	\end{subfigure}
	\qquad
	\begin{subfigure}[t]{0.27\textwidth}
		\centering
		\includegraphics[scale=.25]{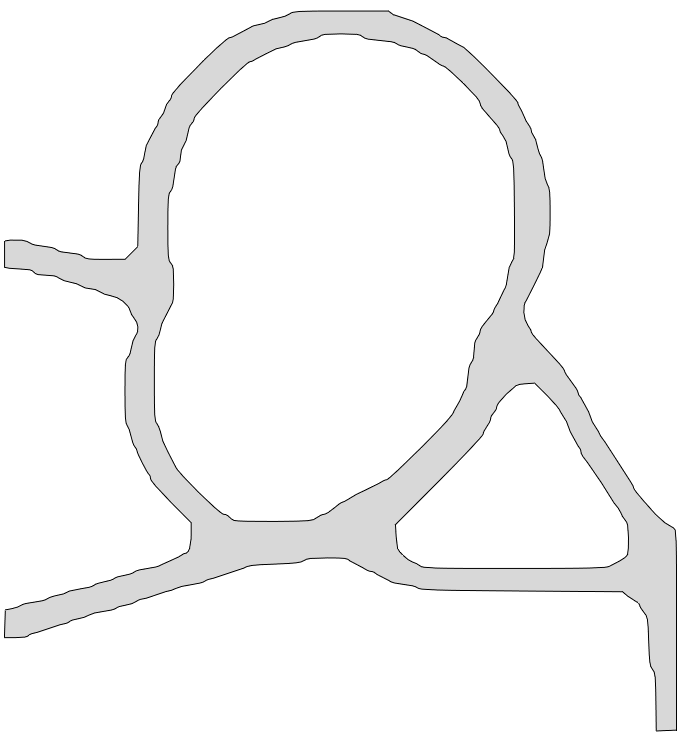}
		\caption{}
		\label{fig:CADR1}
	\end{subfigure}
	\qquad\quad \quad
	\begin{subfigure}[t]{0.27\textwidth}
		\centering
		\includegraphics[scale=.30]{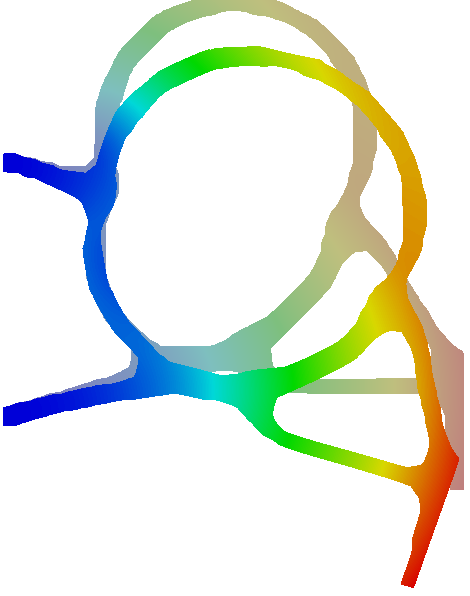}
		\caption{}
		\label{fig:DeforemedR1}
	\end{subfigure}
	\caption{(\subref{fig:actualR1}) The optimized PneuNet intermediate design, (\subref{fig:CADR1}) The CAD model, (\subref{fig:DeforemedR1}) The deformed and undeformed profiles. Red and blue colors indicate maximum and minimum deformation locations. The chamber is pressurized by a 100 bar load.} \label{fig:PneuNet_1}
\end{figure}

 The permitted volume fraction is taken to 0.20. The robust parameter $\Delta \eta = 0.15$ is set. $\beta$ (Eq.~\ref{Eq:projectionfilter}) is varied from 1 to 128, which is doubled at each 50 MMA iterations, and once it reaches its maximum value of 128, it remains so for the rest of the optimization iterations. The dilated volume is updated  every 25 MMA iterations. The higher $\beta$ helps achieve a solution close to 0-1 that is measured herein using the discreteness scale $M_\text{nd}$ defined in \cite{sigmund2007morphology} as
 \begin{equation}
 	M_\text{nd} = \frac{\sum_{e=1}^{{Ne}}4(\bar{\rho_e})(1-\bar{\rho_e})}{{Ne}},
 \end{equation}
   For the mechanism, it is expected that as it is inflated with pressure load, the bottom right corner of it should move down, i.e.,  motion in the negative $y-$direction is sought (Fig.~\ref{fig:DesignDomain}). The flow contrast $\epsilon = \SI{1e-7}{}$ is used. Young's modulus of material $E_1 = \SI{100}{\mega \pascal}$, and Poisson's ration $\nu = 0.40$ are set. The flow parameters $\left\{\eta_\kappa,\,\beta_\kappa\right\} = \left\{0.20,\,10\right\}$ and the drainage parameters $\left\{\eta_d,\,\beta_d\right\} = \left\{0.30,\,10\right\}$ are taken.

The optimized eroded, intermediate, and dilated designs with respective optimum pressure fields are displayed in Fig.~\ref{fig:solution}. The optimized PneuNet design gets an arbitrary-shaped chamber to contain the fluidic pressure load in the optimized designs (Fig.~\ref{fig:solution}), and that is expected to enhance the performance of pneumatic networks. On the other hand, when PneuNets are designed by the heuristic method, they typically have regular-shaped pressure chambers \cite{shepherd2011multigait}. 

The eroded, intermediate, and dilated optimized mechanisms have identical topologies. The intermediate design is taken for the blueprint/fabrication. The eroded designs contain relatively thinner members, whereas the dilated designs are thicker. As per the value of $M_\text{nd}$, the optimized designs are very close to 0-1, i.e., black-white designs are obtained. 
\begin{figure}[h!]
	\begin{subfigure}[t]{0.45\textwidth}
		\centering
		\includegraphics[scale=0.5]{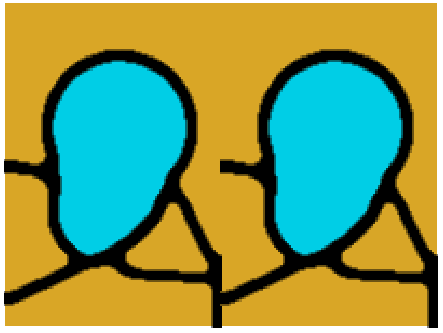}
		\caption{}
		\label{fig:actualR2}
	\end{subfigure}
	\qquad
	\begin{subfigure}[t]{0.45\textwidth}
		\centering
		\includegraphics[scale=.25]{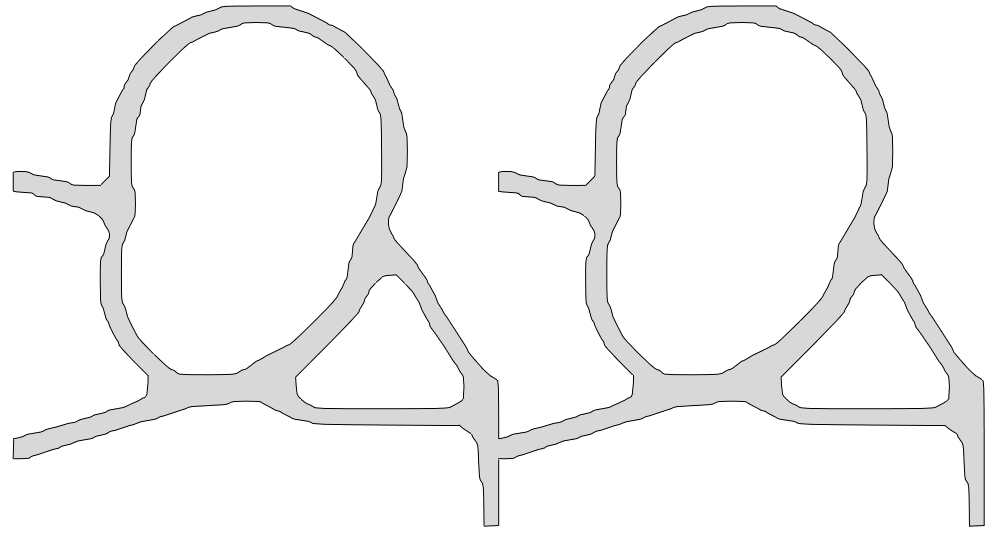}
		\caption{}
		\label{fig:CADR2}
	\end{subfigure}
	\qquad\quad \quad
	\begin{subfigure}[t]{0.90\textwidth}
		\centering
		\includegraphics[scale=.40]{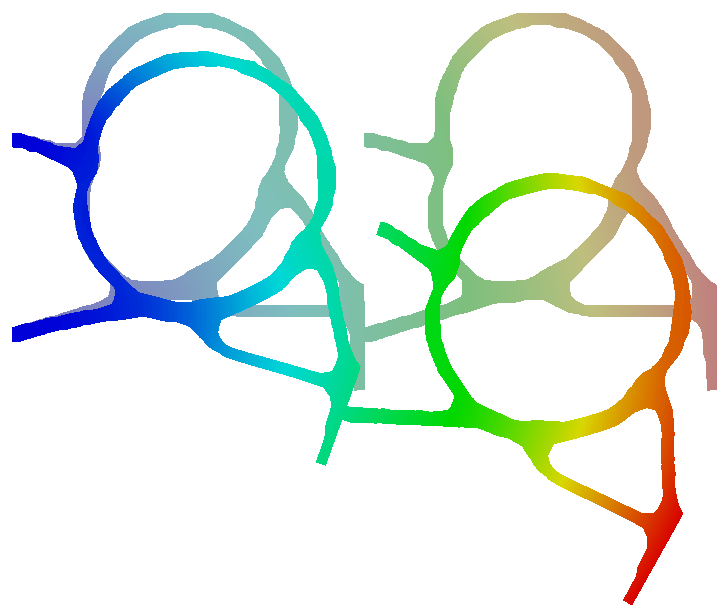}
		\caption{}
		\label{fig:DeforemedR2}
	\end{subfigure}
	\caption{(\subref{fig:actualR2}) The optimized PneuNet  design with two inflating members, (\subref{fig:CADR2}) The CAD model, (\subref{fig:DeforemedR2}) The deformed and undeformed profiles. Each chamber is pressurized by a 100 bar  load.} \label{fig:PneuNet_2}
\end{figure}

\begin{figure}[h!]
	\begin{subfigure}[t]{0.45\textwidth}
		\centering
		\includegraphics[scale=0.5]{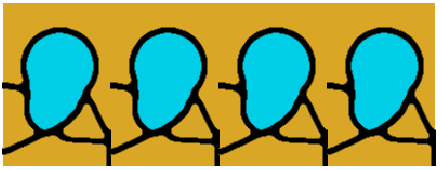}
		\caption{}
		\label{fig:actualR4}
	\end{subfigure}
	\qquad
	\begin{subfigure}[t]{0.45\textwidth}
		\centering
		\includegraphics[scale=.15]{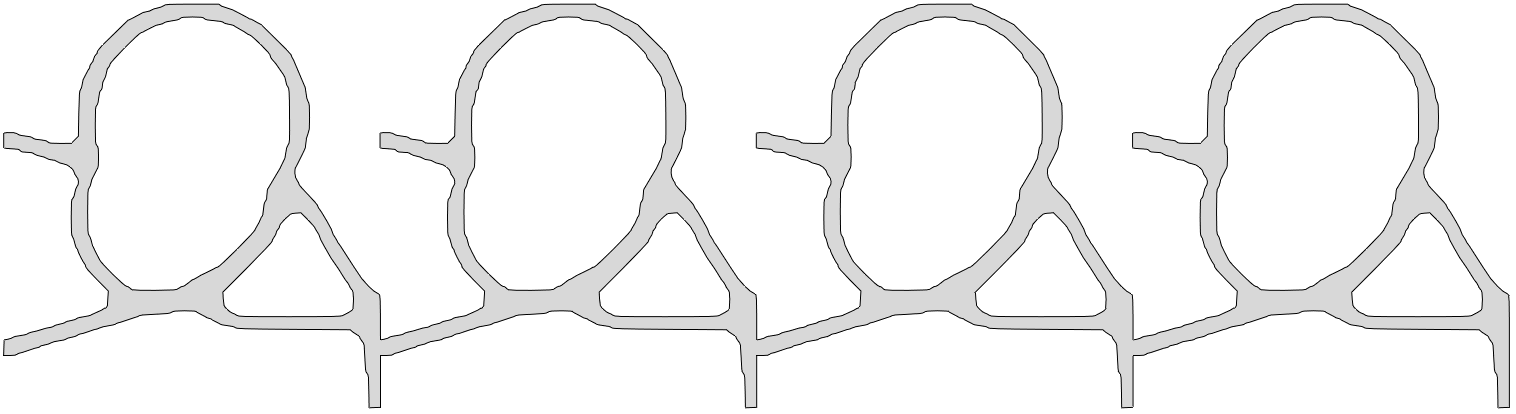}
		\caption{}
		\label{fig:CADR4}
	\end{subfigure}
	\qquad\quad \quad
	\begin{subfigure}[t]{0.90\textwidth}
		\centering
		\includegraphics[scale=.40]{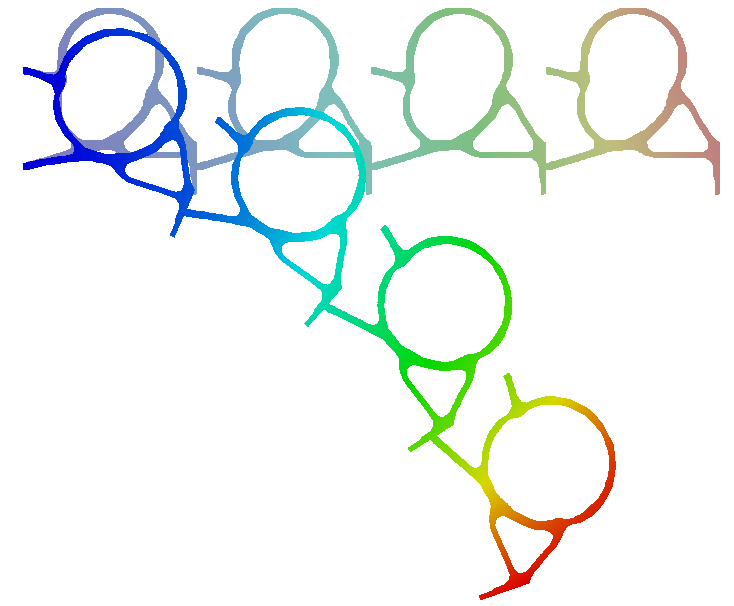}
		\caption{}
		\label{fig:DeforemedR4}
	\end{subfigure}
	\caption{(\subref{fig:actualR2}) The optimized PneuNet  design with four inflating members, (\subref{fig:CADR2}) The CAD model, (\subref{fig:DeforemedR2}) The deformed and undeformed profiles. Each chamber is pressurized by a 100 bar load.} \label{fig:PneuNet_4}
\end{figure}

A member of PneuNet soft robot is designed with $P_\text{in} = \SI{1}{\bar}$ while considering small deformation finite element assumptions in Fig.~\ref{fig:solution}. However, soft robots typically experience finite deformation while performing their tasks. Including nonlinear mechanics formulation within the proposed design approach poses several other challenges, as mentioned in \cite{kumar2022topological}, and is out of the scope of the current study. However, we extract the optimized intermediate design using the technique mentioned in \cite{kumar2022topological}  (Fig.~\ref{fig:intermediatepnet}) and study its behavior with high-pressure loads using a neo-Hookean material model in a commercial software. In addition, we also form different PneuNets from the optimized design displayed in Fig.~\ref{fig:intermediatepnet} and demonstrate their output deformation profiles. In~\cite{caasenbrood2020computational} also a single member is optimized first and combined to get a pneumatic network. They use polygonal/hexagonal~\cite{kumar2022honeytop90} FEs to represent the design domain.

The intermediate optimized design of the soft robot is analyzed in ABAQUS with $P_\text{in} = \SI{100}{\bar}$. The optimized design (Fig.~\ref{fig:actualR1}), its CAD model (Fig.~\ref{fig:CADR1}), and deformed-undeformed profiles  superimposed on each other (Fig.~\ref{fig:DeforemedR1}) are shown in Fig.~\ref{fig:PneuNet_1}. One can note that we get bending motion for the output node with high-pressure loading, which is expected.

Next, we combine two (Fig.~\ref{fig:actualR2}) and four (Fig.~\ref{fig:actualR4}) members of the optimized design to get different PneuNets  (Fig.~\ref{fig:CADR2} and Fig.~\ref{fig:CADR4}). The pressure chambers of the designs are inflated using a $\SI{100}{\bar}$ pressure load. The superimposed deformed and undeformed profiles for these robots are displayed in Fig.~\ref{fig:DeforemedR2} and Fig.~\ref{fig:DeforemedR4}, respectively. As the number of members increases in the network, the final robot deforms relatively more and provides bending motion.  With more pneumatic members and high-pressure loads, the branches may interact and come in contact, i.e., situations for self-contact may occur~\cite{kumar2017implementation,kumar2019computational}  and thus, pose another set of challenges for designing such robots using TO. It can be noted that the pressurized chambers become close to circular shapes in their deformed profiles (Figs.~\ref{fig:DeforemedR1}, \ref{fig:DeforemedR2}, and \ref{fig:DeforemedR4}); this is due because the pressure load acts normal to the surface. In addition, as the number of PneuNet members increases, one gets the complex output motions from the PneuNets, as depicted in Figs.~\ref{fig:DeforemedR1}, \ref{fig:DeforemedR2}, and \ref{fig:DeforemedR4}.

\section{Closure}\label{Sec:Closure}
This paper presents a density-based topology optimization approach to designing soft robots while considering the design-dependent nature of the actuating (fluidic pneumatic) loads. The robust formulation is employed to subdue point/single-node connections. To model the design-dependent character of the pressure load, we use the Darcy law with a conceptualized drainage term per~\cite{kumar2020topology}. The consistent nodal loads are determined from the obtained pressure field using the standard finite element method.

The proposed approach is demonstrated by designing a PneuNet soft robot. A min-max optimization problem is solved wherein the objective is determined using the multi-criteria formulation. The  Method of Moving Asymptotes is used to solve the optimization problem. The optimized PneuNet gives the desired motion and has an arbitrary-shaped pneumatic chamber. The optimized design is combined to form numerous PneuNets, which provide different movements with high-pressure loads. Typically, soft robots experience finite deformation during their performance. Thus, it is necessary to include nonlinear finite element modeling within the optimization formulation, which forms the future direction for this research.

\bibliographystyle{hieeetr}
\bibliography{myreference}

\end{document}